\documentclass[doublecol]{epl2} 
\usepackage[applemac]{inputenc}
\usepackage[T1]{fontenc} 
 \usepackage{amsmath} 
\usepackage{subfigure}
\usepackage{lipsum}
\usepackage{flushend}
\usepackage{amsfonts} 
\usepackage{amssymb, mathrsfs}
\usepackage{braket}
\usepackage{graphicx} 
\usepackage{color}
\usepackage{lipsum}
\usepackage{amsfonts} 
\usepackage{amssymb, mathrsfs}
\usepackage{braket}
\usepackage{bbm}
\def\beq{\begin{equation}}
\def\eeq{\end{equation}}
\def\bsp{\begin{split}}
\def\esp{\end{split}}
\def\bea{\begin{eqnarray}}
\def\eea{\end{eqnarray}}
\def\ba{\begin{array}}
\def\ea{\end{array}}

\def\dg{\dagger}

\def\lb{\left(}
\def\rb{\right)}

\def\l.{\left.}
\def\r.{\right.}

\def\ra{\rangle}
\def\la{\langle}

\def\bo{{\bf k}}

\title{Topological  magnon nodal-lines and absence of magnon spin Nernst effect in layered collinear antiferromagnets}
\shorttitle{Topological nodal-line magnonic phase and absence of magnon spin Nernst effect} 

\author{S. A. OWERRE\inst{1} }

\shortauthor{S. A. OWERRE}

\institute{                   
  \inst{1} Perimeter Institute for Theoretical Physics - 31 Caroline St. N., Waterloo, Ontario N2L 2Y5, Canada\\
}

 {Nonelectronic thermal conduction and heat-pulse propagation in solids; thermal waves}

\abstract{
We propose the existence of a symmetry-protected  topological Dirac nodal line (DNL) magnonic phase in layered honeycomb collinear antiferromagnets even in the presence of spin-orbit  Dzyaloshinskii-Moriya interaction. We show that the magnon spin Nernst effect, predicted to occur in strictly two-dimensional (2D) honeycomb collinear antiferromagnets  cancels out in the layered honeycomb collinear antiferromagnets. In other words, the magnon spin Nernst effect in  each 2D antiferromagnetic layer cancels out the succeeding layer. Hence, the Berry curvature vanish in the entire Brillouin zone due to the combination of time-reversal and space-inversion ($\mathcal{PT}$) symmetry.    However,  upon  symmetry breaking by an external magnetic field, we show that a non-vanishing Berry curvature and Chern number protected topological magnon bands are induced in the non-collinear spin structure. This leads to an experimentally accessible magnon thermal Hall effect  in the $\mathcal{PT}$ symmetry-broken topological DNL magnonic phase of layered honeycomb antiferromagnets. We propose that the current predicted results can be experimentally investigated in the layered honeycomb  antiferromagnets CaMn$_2$Sb$_2$, BaNi$_2$V$_2$O$_8$, and  Bi$_3$Mn$_4$O$_{12}$(NO$_3$).
}

\begin{document}

\maketitle

\section{Introduction}
 The theoretical predictions and experimental discoveries of electronic topological semimetals  \cite{wan,bur,xu,lv, wan1, wan2, tang} have attracted considerable attention in condensed-matter physics. This has prompted the re-examination of topological band structures  in solid-state materials. Essentially, the   condensed-matter realization of topological semimetals with linear band crossing points is independent of the statistical nature of the quasiparticle excitations.  Therefore, it can be  extended to bosonic quasiparticle excitations \cite{lu,fee}. In particular, the magnonic analogs of electronic topological semimetals, featuring linear magnon band crossing at finite energy, are currently one of the active research areas in quantum magnetism \cite{mw1, mw2, mw3, mw4, mw5, mw6, mw7, mw8,mw10,mw11,mw12,mw13,mwc}. Three-dimensional Dirac nodal-line magnons with linear band crossings are predicted to occur in  3D collinear ferromagnets  \cite{mw12, mw4} when the spin-orbit Dzyaloshinskii-Moriya (DM) interaction \cite{dm,dm2} is ignored. They are immediately gapped out in the presence of a nonzero DM interaction, and thus are similar to electronic nodal-line semimetals \cite{NL0,NL1, NL2, NL3, NL4}. Recently,  symmetry-protected electronic topological nodal-line semimetals which are robust in the presence of spin-orbit coupling  have been proposed and experimentally observed \cite{NL5}.
 
 Moreover, recent theoretical work has predicted that the DM interaction can induce topological  DNL magnons in the 3D collinear antiferromagnet Cu$_3$TeO$_6$ \cite{kli}. However, the spin canting induced by the DM interaction was found to be very negligible, hence  no evidence of DNL magnons was observed in  recent inelastic neutron scattering  experiments on  Cu$_3$TeO$_6$ \cite{yao,bao}. Thus, 3D topological DNL magnonic phase has not yet been found in real magnetically ordered materials.  In addition, the antiferromagnetic material Cu$_3$TeO$_6$ shows very complex structure. For instance,  the bulk material has 12 sites in the unit cell and up to ninth nearest-neighbour interactions \cite{yao,bao}, which are very difficult to comprehend analytically.  Therefore, we seek for a simpler  3D collinear antiferromagnetic system  in which the 3D topological DNL magnonic phase predicted in the collinear antiferromagnet Cu$_3$TeO$_6$ can be clearly manifested.

In this letter, we present three new results. First, we show that layered honeycomb collinear antiferromagnetic materials are symmetry-protected  3D topological DNL magnonic systems.  The  topological DNL magnons in these systems \cite{note1} are robust against the  DM interaction, and are  protected by magnetic crystal symmetry. They are analogous to predicted DNL magnons in Cu$_3$TeO$_6$ \cite{kli}.  The major differences in the current model are as follows: (1) The current model applies to a variety of layered honeycomb antiferromagnetic materials, which are endowed with only two magnon branches. Therefore, the topological DNL magnons can be clearly seen at the lowest acoustic magnon branch.   (2)  The topological DNL magnons in the current system are not induced by the DM interaction due to the inherent structure of the honeycomb lattice. (3) The topological DNL magnons do not require further nearest-neighbour interactions apart from the first one. In addition, we show that the present topological DNL magnons show small dispersive (nearly flat) drumhead surface states with DNL rings projected on the (001) surface Brillouin zone (BZ). 

Second, we show that the layered honeycomb collinear antiferromagets behave quite differently from the 2D honeycomb collinear antiferromagnets  in that $\mathcal{PT}$ symmetry forces the Berry curvature to vanish in the entire BZ.  Thus, the magnon spin Nernst effect predicted to occur in 2D honeycomb collinear antiferromagnets \cite{ran,kov,hoon} is prohibited in the layered honeycomb collinear antiferromagnetic materials. We note that a recent experiment did not successfully observe any clear discernible magnon spin Nernst voltage in the bulk layered honeycomb collinear antiferromagnet MnPS$_3$ \cite{shi} at zero magnetic field. Therefore, our result may explain the absence of magnon spin Nernst voltage in the layered honeycomb collinear antiferromagnets.

Third, we show that upon breaking of symmetry by an external magnetic field, a topological phase transition occurs from the 3D topological DNL magnons to a 3D magnon Chern insulator.  We compute the Berry curvature, Chern numbers, and chiral magnon edge modes of the 3D antiferromagnetic magnon Chern insulator. We also predict that the experimentally accessible magnon thermal Hall effect \cite{th1,th2,th3, th4, th5} will occur in  the $\mathcal{PT}$ symmetry-broken layered honeycomb antiferromagnets.  We believe that the current predictions are  pertinent to experimental investigation in the bulk layered honeycomb antiferromagnetic materials  such as CaMn$_2$Sb$_2$ \cite{mcn}, BaNi$_2$V$_2$O$_8$ \cite{kly},  Bi$_3$Mn$_4$O$_{12}$(NO$_3$)\cite{matt}, and others \cite{matt,oku,doi,shia, shi1, zhou, will}.

\section{Heisenberg spin  model}
 We consider the microscopic spin Hamiltonian of layered honeycomb-lattice antiferromagnetic systems in the presnece of  DM interaction and an external  Zeeman magnetic field. The model is  governed by 
\begin{align}
\mathcal H&=J\sum_{ \la ij\ra,\ell} {\bf S}_{i,\ell}\cdot{\bf S}_{j, \ell}+\sum_{ \la\la ij\ra\ra,\ell}{\bf D}_{ij,\ell}\cdot{\bf S}_{i,\ell}\times{\bf S}_{j, \ell}\nonumber\\&+J_c\sum_{\la \ell \ell^\prime\ra, i} {\bf S}_{i,\ell}\cdot{\bf S}_{i,\ell^\prime}-{\bf H}\cdot\sum_{i,\ell}{\bf S}_{i,\ell},
\label{model}
\end{align}
where ${\bf S}_{i,\ell}$ is the spin vector at site $i$ in layer ${\ell}$. The first term is   the nearest-neighbour (NN) intralayer antiferromagnetic coupling. The second term is the out-of-plane next-nearest-neighbour (NNN) DM interaction ${\bf D}_{ij,\ell}= \nu_{ij,\ell}D {\bf \hat z}$, which is  allowed on the honeycomb lattice due to inversion symmetry breaking with $\nu_{ij,\ell}=\pm 1$. The third term is the NN interlayer antiferromagnetic coupling.  Finally, the last term is an external Zeeman magnetic field along the stacking direction ${\bf H}= g\mu_B H{\bf \hat z}$, where $g$ is the spin g-factor and $\mu_B$ is the Bohr magneton. In Fig.~\ref{lattice}(a) and (b) we have shown the top view of the  honeycomb-lattice antiferromagnet stacked congruently  along the (001) direction and its bulk Brillouin zone (BZ) respectively. 

\section{Topological nodal-line magnons in the collinear spin structure}

\subsection{Symmetry protections of  collinear spin structure}
At zero magnetic field $H=0$, the ground state of the Hamiltonian is a collinear N\'eel ordered state with zero net magnetization. In this case, the collinear N\'eel magnetization can point in any direction, but the DM interaction has to be parallel to the magnetization direction in order to have a significant effect in the low-lying noninteracting magnon excitations. Here, we assume that the collinear N\'eel magnetization and the DM vector point along the $z$-direction. This collinear spin state is invariant under  $\mathcal{PT}$ symmetry since $\mathcal{P}$ symmetry interchanges the layers and $\mathcal{T}$ symmetry flips the spins on each layer. In addition, the system also possesses $\mathcal M_z$ mirror reflection symmetry about the $x$-$y$ plane that sends $z\to-z$. For ferromagnetic interlayer coupling $J_c<0$, there is only one single doubly-degenerate magnon band. However,  for antiferromagnetic  interlayer coupling $J_c>0$, there are two  non-degenerate magnon branches as confirmed experimentally in the layered honeycomb antiferromagnetic materials CaMn$_2$Sb$_2$ \cite{mcn} and BaNi$_2$V$_2$O$_8$ \cite{kly}.    We do not consider the  ferromagnetic interlayer coupling as it is similar to 2D honeycomb antiferromagnets \cite{ran,kov}, which is topologically trivial. The two non-degenerate magnon branches in  the case of antiferromagnetic  interlayer coupling give the possibility for investigating  topological linear band crossing in this system.

 \begin{figure}
\centering
\includegraphics[width=0.9\linewidth]{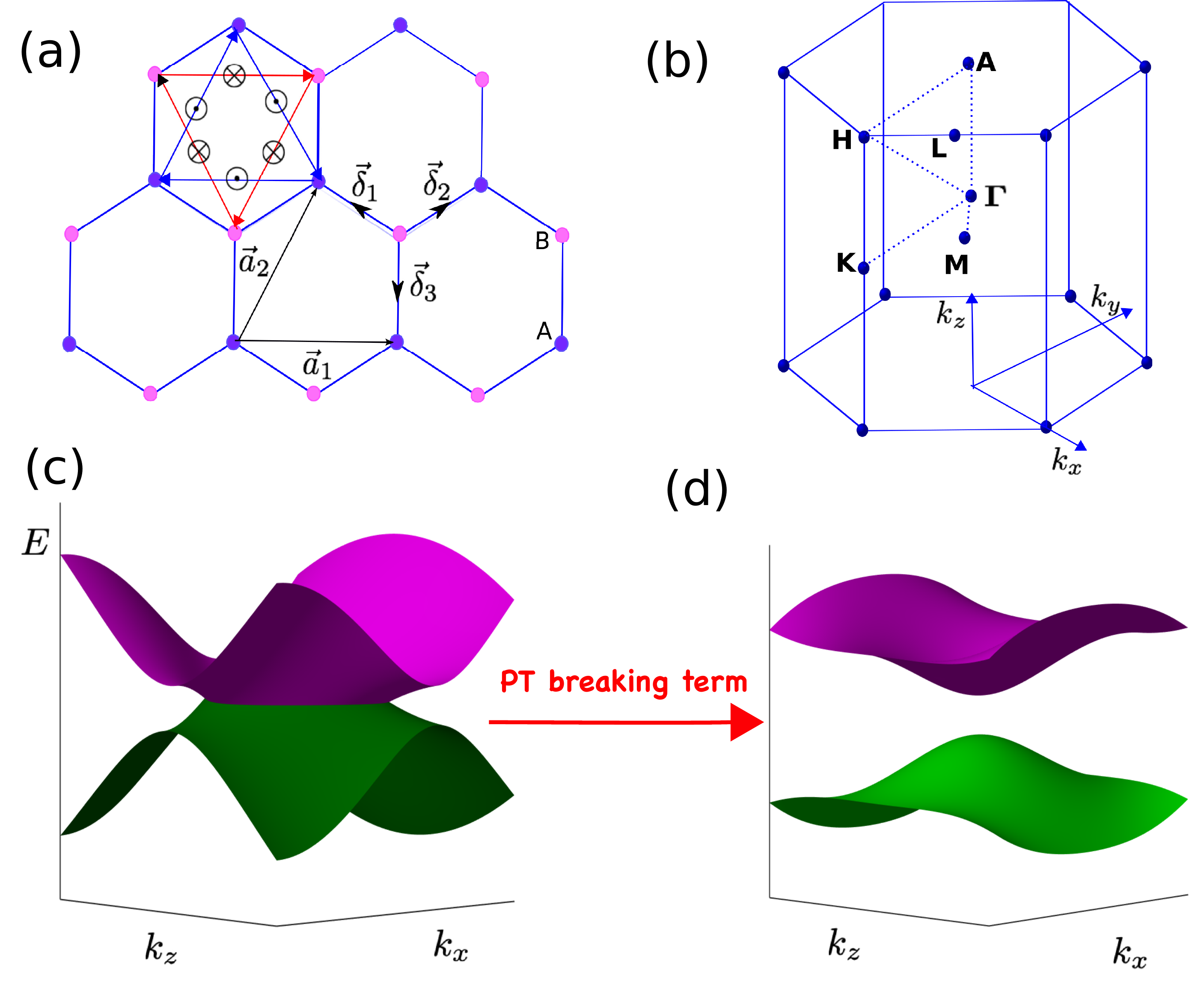}
\caption{Color online. (a) Top view of honeycomb antiferromagnets stacked congruently along the (001) direction.  The crossed and dotted circles denote the alternating DM interaction along the NNN bonds.  $A$ and $B$ denote the two sublattices of the honeycomb lattice. (b)  3D bulk Brillouin zone of the hexagonal lattice. (c) Symmetry-protected DNL magnons on the $k_y=0$ plane. (d) Magnon Chern insulator induced by symmetry-broken topological DNL magnons.}
\label{lattice}
\end{figure}

\begin{figure}
\centering
\includegraphics[width=0.9\linewidth]{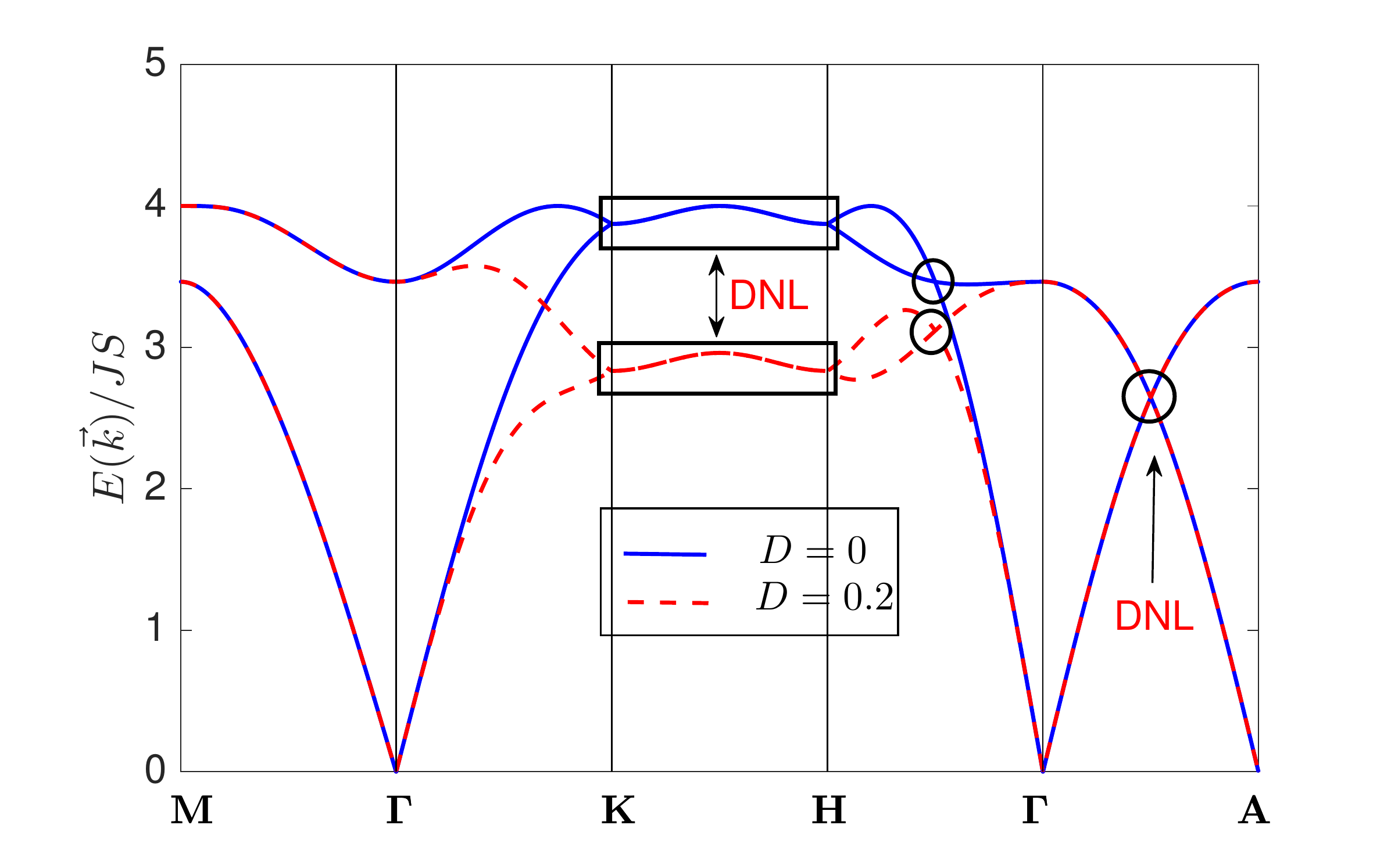}
\caption{Color online. Antiferromagnetic Dirac nodal line magnon bands for $J_c/J=0.5$ and $H/J=0$. The DNL magnons are indicated by black rectangles and circles.}
\label{DNL}
\end{figure}


\subsection{Bosonic Hamiltonian} 

The magnon dispersions in the low-temperature regime can be captured by the Holstein Primakoff (HP)  transformation \cite{hp}.  We define the HP transformation for each layer $\ell$ as  
\begin{align}
S_{i,\ell}^{ z}= S-a_{i,\ell}^\dagger a_{i,\ell},~S_{i,\ell}^+\approx \sqrt{2S}a_{i,\ell}=(S_{i,\ell}^-)^\dg,
\end{align}
for $i\in  \rm{sublattice }~ A$ and 
\begin{align}
S_{j,\ell}^{ z}= -S+a_{j,\ell}^\dagger a_{j,\ell},~S_{j,\ell}^+\approx \sqrt{2S}a_{j,\ell}^\dg=(S_{j,\ell}^-)^\dg
\end{align}
for $j\in \rm{sublattice }~B$, where $a_{i,\ell}^\dagger (a_{i,\ell})$ are the bosonic creation (annihilation) operators and  $S^\pm_{i,\ell}= S^x_{i,\ell} \pm i S^y_{i,\ell}$ denote the spin raising and lowering  operators. We note that the sublattice $A$ on layer $\ell$ couples antiferromagnetically to sublattice $A$ on layers  $\ell^\prime$. A similar argument applies  to the sublattice $B$.  Hence,  the HP transformation  has the same form for layers $\ell$ and $\ell^\prime$.   

The resulting non-interacting magnon Hamiltonian in momentum space is given by  $\mathcal H=\frac{1}{2}\sum_{{\bf k}} \psi^\dg({{\bf k}})\cdot \mathcal H({{\bf k}})\cdot \psi({{\bf k}}),$
where $\psi^\dg({{\bf k}})=(a_{{\bf k},A}^\dg,a_{{\bf k},B}^\dg, a_{-{\bf k},A},a_{-{\bf k},B})$ is the basis vector. 
\begin{align}
 \mathcal H({{\bf k}})=
 \begin{pmatrix}
 \mathcal{A}({\bf k})&\mathcal{B}({\bf k})\\ \mathcal{B}^*(-{\bf k})&\mathcal{A}^*(-{\bf k})
 \end{pmatrix},
 \label{hamp1}
\end{align}
where
\begin{align}
 \mathcal{A}({{\bf k}})&=\big[f_0+f^{D}({\bf k_\parallel})\big] {\rm {\bf I}}_{2\times 2},
 \\ \mathcal{B}({{\bf k}})&=
 \begin{pmatrix}
f(k_z)&f({\bf k_\parallel})\\ f^*({\bf k_\parallel})&f(k_z)
 \end{pmatrix},
\end{align}
where ${\rm {\bf I}}_{2\times 2}$ is an identity $2\times 2$ matrix, $f_0=3JS+2J_cS$, $f(k_z)=2J_cS\cos k_z$,  $f({\bf k_\parallel})=JS\lb 1 + e^{-ik_{1\parallel}}+e^{-ik_{2\parallel}}\rb$, and  $f^{D}({\bf k_\parallel})=-f^{D}(-{\bf k_\parallel})=2DS\lb\sin k_{1\parallel} -\sin k_{2\parallel}+\sin (k_{2\parallel}-k_{1\parallel})\rb$. The in-plane momentum is given by  $k_{i\parallel}={\bf k_\parallel}\cdot{ \bf a}_i$ and the primitive vectors of the honeycomb lattice are  ${ \bf a}_1=\sqrt{3}\hat x$  and ${ \bf a}_2= \sqrt{3}\hat x/2 + 3\hat y/2$.  The 3D momentum vector is ${\bf k}=({\bf k_\parallel},k_z)$, where  $ {\bf k_\parallel}=(k_x,k_y)$ is the 2D in-plane wave vector.

\subsection{Topological Dirac nodal-line magnons  and drumhead surface states } 

The Hamiltonian \eqref{hamp1} can be diagonalized by the  paraunitary operator $\mathcal U_\bo$ (see Supplemental Material).  The resulting magnon energy branches  are given by  \cite{note1}  
 \begin{align}
  E_{\pm}({\bf k})=f^{D}({\bf k_\parallel})+\sqrt{f_0^2-\big[f(k_z)\pm |f({\bf k_\parallel})|\big]^2}.
  \label{band}
 \end{align}
 The condition for topological  DNL magnons to exist in this system is given by
 \begin{align}
 f( k_z)f({\bf k_\parallel})=0,
 \end{align}
which is satisfied when $f( k_z)=0$ or $f({\bf k_\parallel})=0$. The former gives $k_z=\pi/2$ for fixed ${\bf k_\parallel}$, and the latter gives ${\bf k_\parallel}=\bar{\bf{K}}=(\pm 4\pi/3\sqrt{3}, 0)$ for fixed $k_z$. 

In Fig.~\eqref{DNL} we have shown the DNL magnon bands along the high-symmetry lines of the BZ for zero and non-zero DM interaction.  In stark contrast to 3D collinear ferromagnets for which  the DNL magnons  transform to Weyl magnons \cite{mw2, mw12, mw3, mw4} upon the inclusion of the DM interaction, we find that the antiferromagnetic DNL magnons in the collinear honeycomb antiferromagnets  persist even in the presence of  the DM interaction as shown in Fig.~\ref{DNL}(b). The DM interaction  introduces asymmetry between the DNL magnons centred at  $\pm\bar{\bf {K}}$  in the fixed $k_z$ plane, with $f^{D}(\pm\bar{\bf{K}})=\mp 3\sqrt{3}D$. 


 \begin{figure*}
\centering
\includegraphics[width=0.9\linewidth]{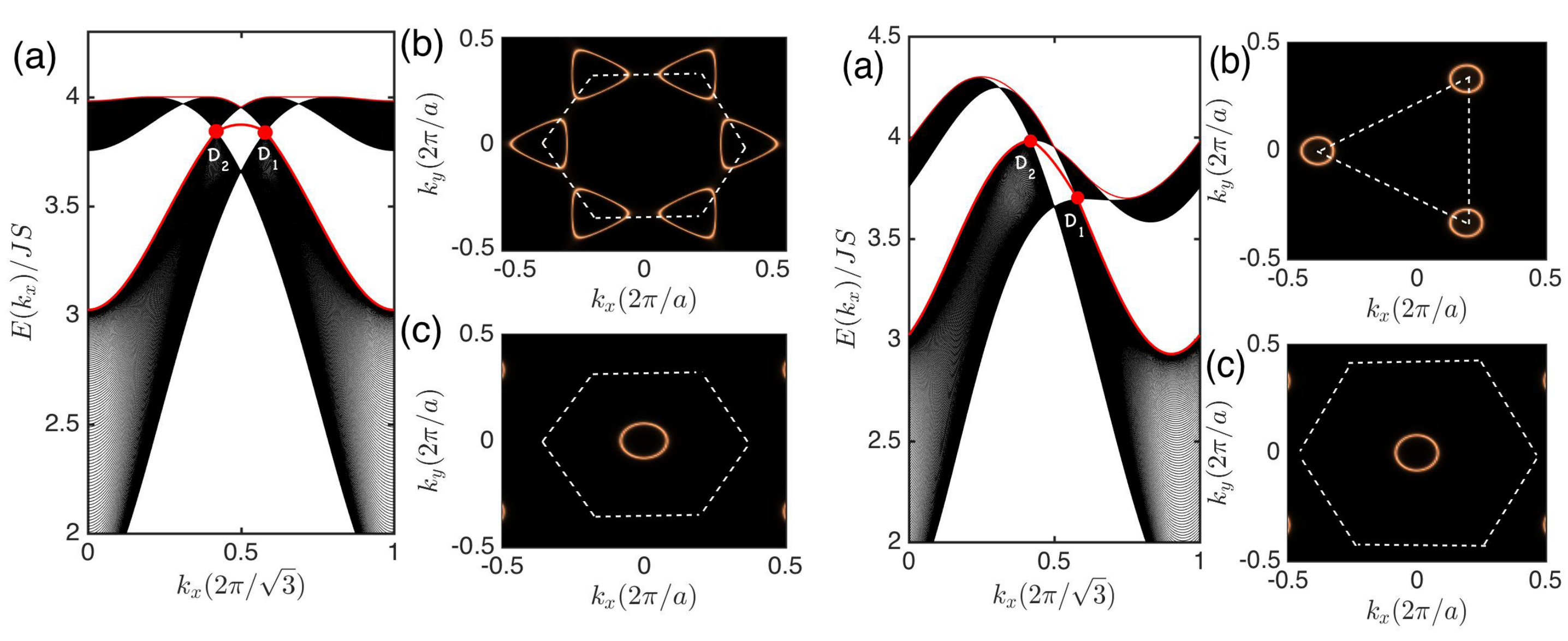}
\caption{Color online. The projected (010)-surface dispersion along $k_x$ direction on the $k_z=0$ plane and the Dirac magnon rings (yellowish rings) on the surface BZ for  $H/J=0$ and $J_c/J=0.5$.  Left panels: $D/J=0$.   Right panels: $D/J=0.2$. The dashed hexagon indicates  the  (001) surface BZ. The energy is set at the Dirac nodal line at the ${\bf K}$ point (b) and along the ${\bf \Gamma}$--${\bf A}$ line (c). }
\label{DNL_loop}
\end{figure*}

One of the advantages of the present model over other spin models \cite{mw12,kli} is that the analysis for DNL magnons can almost be done analytically. To obtain the expression for the DNL magnons, we consider the fixed $k_z=k_z^0$ plane.  The equation for the DNL magnons is given by $E_\pm({\bf k_\parallel} ,k_z^0)=E_{\text{DNL}}$. Expanding near ${\bf k_\parallel}=\bar{\bf \bf {K}}=(\pm 4\pi/3\sqrt{3}, 0)$ the solution to this equation yields
\begin{align}
v_s^2(q_x^2+q_y^2)=f^2(k_z^0),
\end{align}
where  ${\bf q}={\bf k_\parallel}-\bar{\bf{K}}$ and $v_s=3JS/2$. This  gives a Dirac nodal magnon ring of radius $r_1=|f(k_z^0)/v_s|$ as shown in Fig.~\ref{lattice}(c).  Similarly, expanding near ${\bf k_\parallel}=\bar{\boldsymbol  \Gamma}=(0, 0)$ yields
\begin{align}
v_0^2(q_x^2+q_y^2)=2v_0^2-f^2(k_z^0),
\end{align}
where  ${\bf q}={\bf k_\parallel}-\bar{\boldsymbol \Gamma}$, and $v_0=3JS/\sqrt {2}$. This also gives a  DNL ring of radius $r_2=\sqrt{|2v_0^2-f^2(k_z^0)|}/v_0$. In both cases, the DNL rings are independent of the DM interaction and  they are present provided the interlayer coupling does not vanish. This means that both strongly-coupled  and weakly-coupled honeycomb antiferromagnetic layers are candidates for topological DNL magnons.

One of the hallmarks of DNLs is the drumhead surface states that connect projected DNLs on the surface BZ \cite{NL1, NL2, NL3, NL4, NL5}. In the present case, the system shows small dispersive (nearly flat) drumhead surface states connecting the (010) surface projection of the DNL magnons which form Dirac nodal rings in the surface BZ as shown in the  Fig.~\ref{DNL_loop}  at zero DM interaction (left) and nonzero DM interaction (right).  We can explicitly see that the Dirac magnon rings are centred at $\pm\bar{{\bf K}}$ for the DNL along ${\bf K}$-- ${\bf H}$ line (b) and at $\bar{\bf \Gamma}$  for the DNL along ${\bf \Gamma}$--${\bf A}$ line. The effect of the DM interaction is to make the DNL asymmetric with different energies, however the nearly flat drumhead surface states still persist as shown in Fig.~\ref{DNL_loop} (right).
\begin{figure}
\centering
\includegraphics[width=1\linewidth]{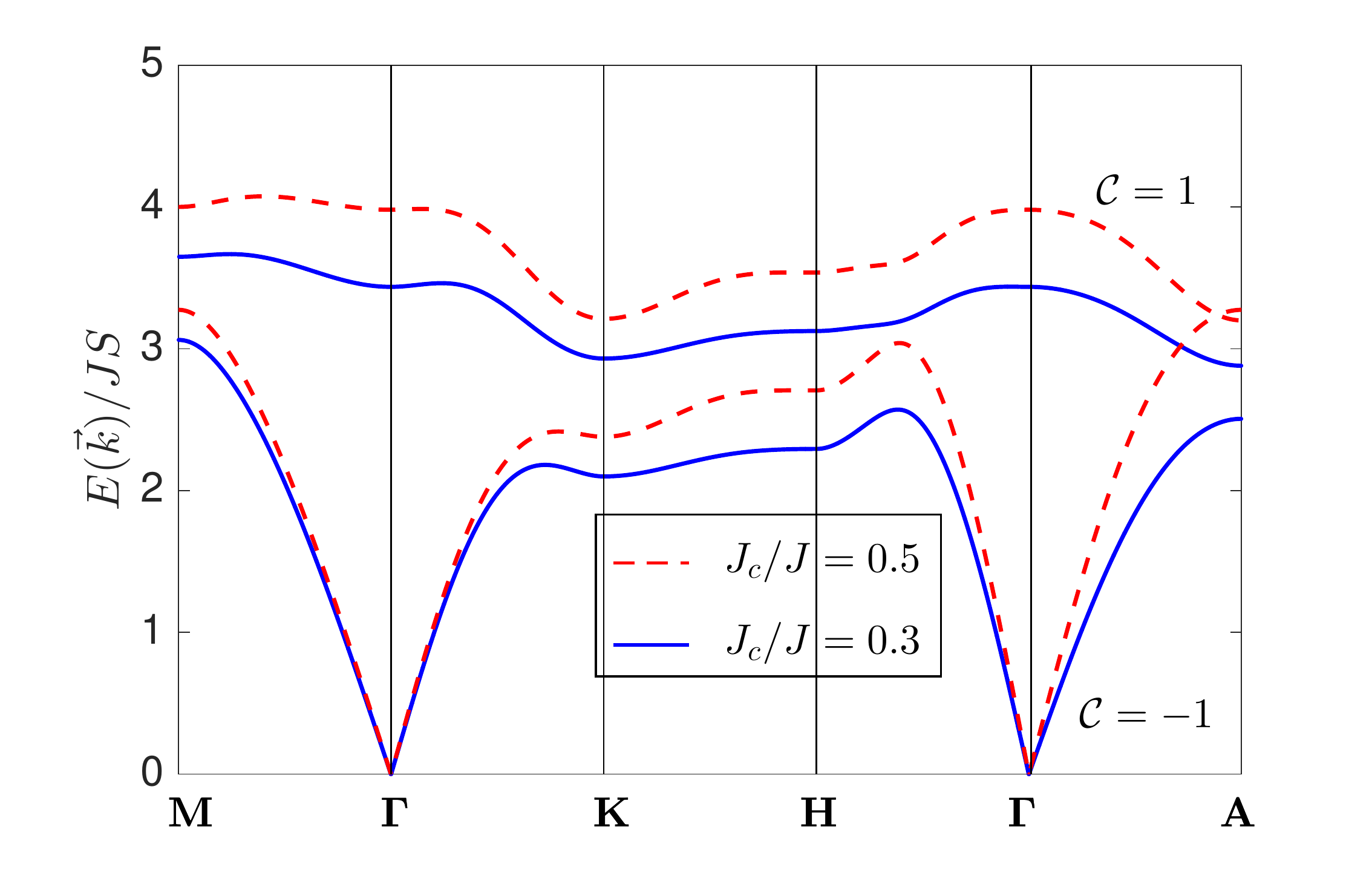}
\caption{Color online. Antiferromagnetic magnon Chern insulator bands for $D/J=0.2$ and $H=0.3H_S$. The lower and upper magnon bands carry Chern numbers $\mathcal C=\mp 1$ respectively.}
\label{MCI}
\end{figure}

\section{Absence of magnon  spin Nernst effect}

In the single-layer honeycomb collinear antiferromagnet $J_c =0$,  the presence of DM interaction induces a nonzero Berry curvature. Due to time-reversal symmetry, the thermal Hall response vanishes, whereas the  magnon spin Nernst response is nonzero \cite{ran,kov,hoon}.   For the layered honeycomb collinear antiferromagnets coupled antiferromagnetically {\it i.e} $J_c >0$, we showed above  that  the magnon branches are not doubly-degenerate as also confirmed experimentally in CaMn$_2$Sb$_2$ \cite{mcn} and BaNi$_2$V$_2$O$_8$ \cite{kly}.  In order to investigate the magnon transport in this system, let us define the Berry curvature of the magnon bands  as \cite{th5}
\begin{align}
 [\Omega_{\alpha\beta}(\bo)]_{nn}=-2\mathfrak{I}[\tau_3\mathcal (\partial_{k_\alpha}\mathcal U_\bo^\dg)\tau_3(\partial_{k_\beta}\mathcal U_\bo)]_{nn},
 \label{bc1}
 \end{align}
where $\alpha,\beta=x,y,z$. $\mathfrak{I}$ denotes  imaginary part, and $\tau_3$ is defined in the Supplemental Material. We find that the components of the Berry curvature are given by 
\begin{align}
[\Omega_{\alpha\beta}(\bo)]_{11}&=[\Omega_{\alpha\beta}(\bo)]_{33}\nonumber\\&=-\mathfrak{I}\Big[\partial_{k_\alpha}\big(e^{-i\varphi_{{\bf k_\parallel}}}\cosh\phi_{\bo}\big)\partial_{k_\beta}\big(e^{i\varphi_{{\bf k_\parallel}}}\cosh\phi_{\bo}\big)\nonumber\\&-
\partial_{k_\alpha}\big(e^{-i\varphi_{{\bf k_\parallel}}}\sinh\phi_{\bo}\big)\partial_{k_\beta}\big(e^{i\varphi_{{\bf k_\parallel}}}\sinh\phi_{\bo}\big)\Big].
\end{align}
Similarly, 
\begin{align}
[\Omega_{\alpha\beta}(\bo)]_{22}=[\Omega_{\alpha\beta}(\bo)]_{44}=-\mathfrak{I}\big[\phi_{\bo}\to\theta_{\bo}\big],
\end{align}
where $\phi_{\bo}, \theta_{\bo}, \varphi_{\bf k_\parallel}$ are defined in the Supplemental Material. 
Evidently, we can see that the Berry curvature vanishes identically 
\bea 
[\Omega_{\alpha\beta}(\bo)]_{nn}=0.
\eea
This means that both the magnon thermal Hall response $(\kappa_{xy})$ \cite{th1,th2,th3,th4,th5} and the magnon  spin Nernst response $(\alpha_{xy}^s)$ \cite{ran,kov}  vanish as well. Therefore, layered honeycomb collinear antiferromagnets coupled antiferromagnetically do not possess any discernible $\kappa_{xy}$  and $\alpha_{xy}^s$.   This result may explain the reason why no clear  magnon spin Nernst voltage was observed in the bulk honeycomb antiferromagnet MnPS$_3$ at zero magnetic field \cite{shi}.  Therefore, layered honeycomb antiferromagnets are symmetry-protected topological DNL magnonic systems.

The topological protection of the DNL magnons can be inferred from the  Berry phase defined as 
\bea
\gamma_n=\oint_{ C} \mathcal F_n(\bo)\cdot d{ \bo},
\eea
over a closed loop $ C$, where $\mathcal F_n(\bo)$ is the Berry connection given by \bea \mathcal F_n(\bo)=-i\big[\braket{\tau_3\mathcal U_{\bo}^\dg|{\vec \nabla}_{\bo}\tau_3\mathcal U_{\bo}}\big]_{nn}.\eea 
For a closed path encircling the topological DNL magnons in momentum space, we find that the Berry phase is $\gamma_n=\pm\pi$, otherwise $\gamma_n=0$. Therefore the topological DNL magnons can be interpreted as topological defects in 3D momentum space. A similar result can be obtained by considering the eigenvalues of the mirror reflection symmetry $\mathcal M_z$ \cite{NL5}.    

\begin{figure}
\centering
\includegraphics[width=1\linewidth]{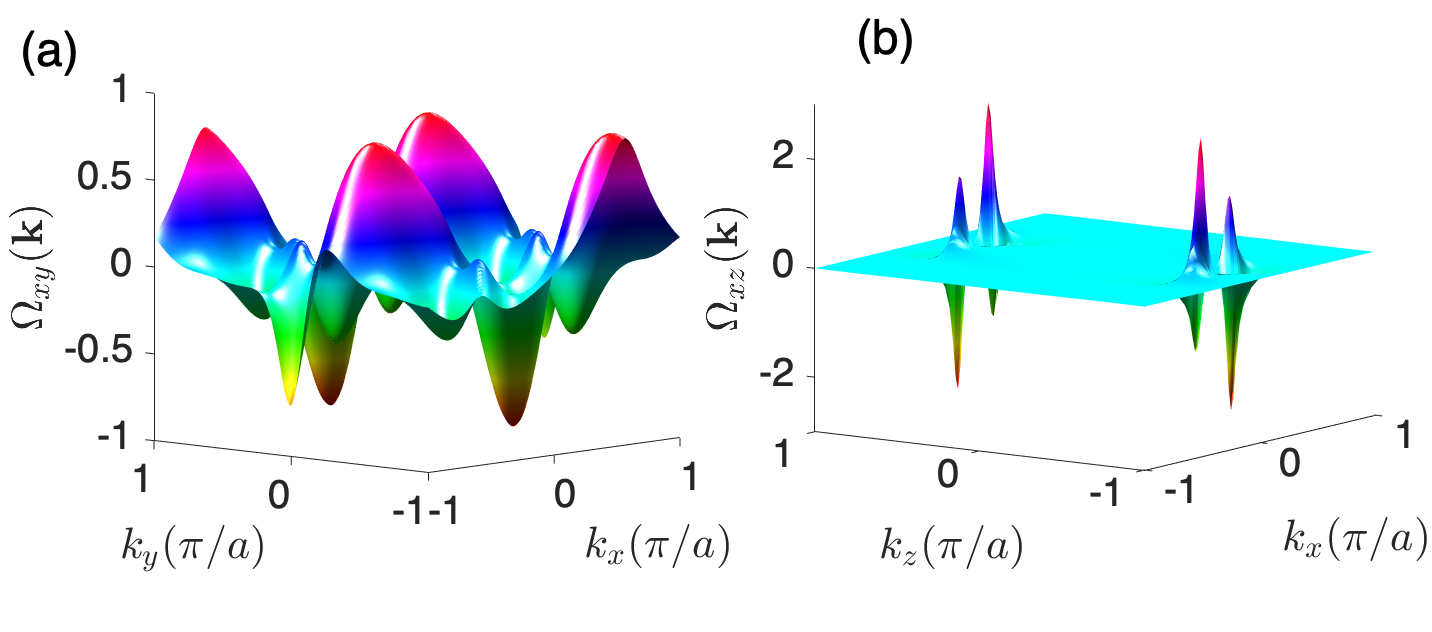}
\caption{Color online. The Berry curvatures  $\Omega^n_{xy}({\bf k})$ and  $\Omega^n_{xz}({\bf k})\equiv \Omega^n_{yz}({\bf k})$ of the lowest topological magnon band $(n=1)$  on the $k_z =0$ plane (a) and on the $k_y =0$ plane (b).  The parameters for both panels are $D/J=0.2$, $J_c/J=0.3$ and $H=0.3H_S$.}
\label{BC}
\end{figure}

\begin{figure}
\centering
\includegraphics[width=1\linewidth]{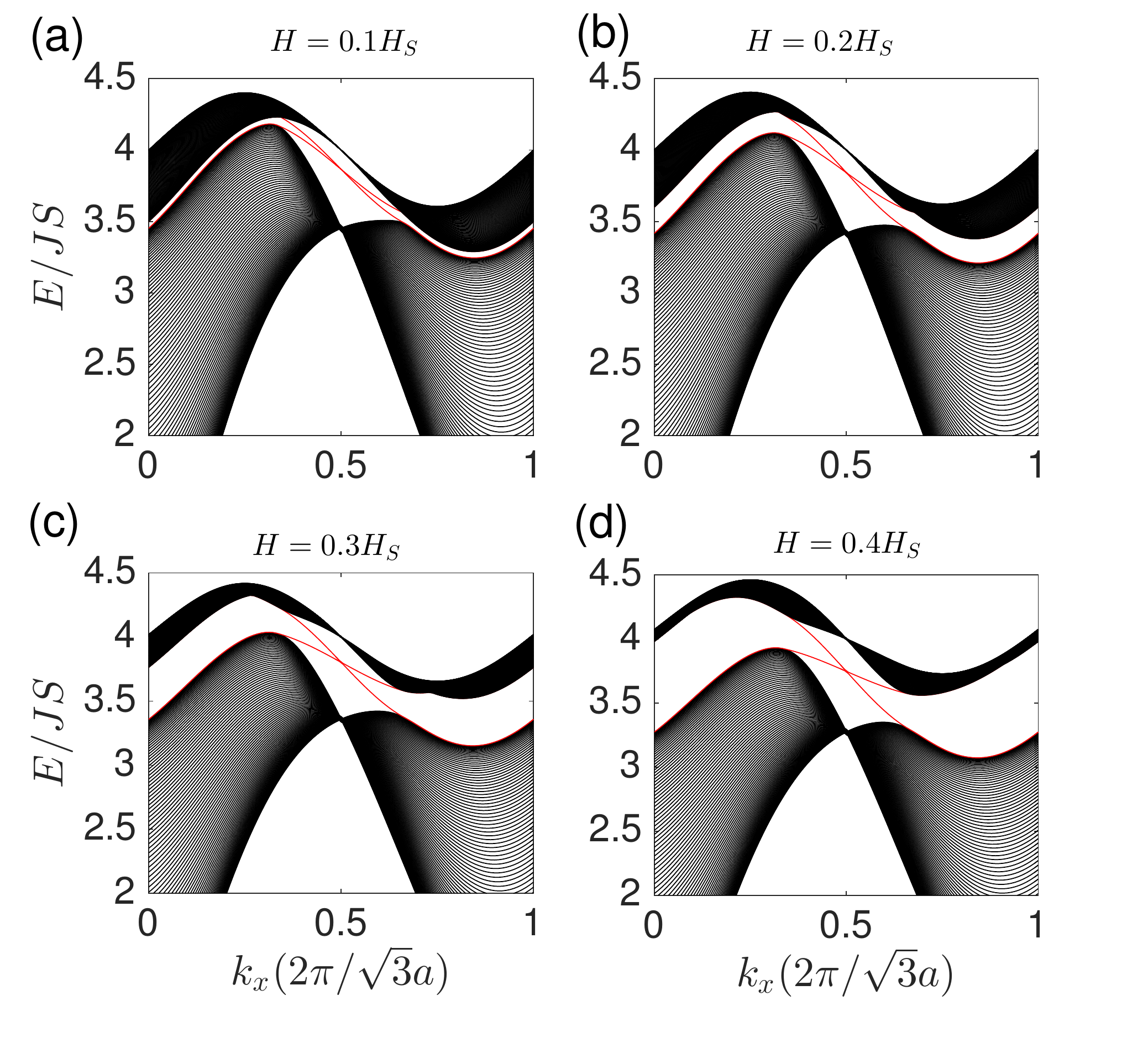}
\caption{Color online.  Evolution of the projected (010)-surface dispersion and chiral magnon edge modes denoted by red curves along the $k_z=0$ line for different values of  magnetic field.   The parameters are  $D/J=0.2$ and $J_c/J=0.5$. }
\label{Edge}
\end{figure}

\begin{figure}
\centering
\includegraphics[width=1\linewidth]{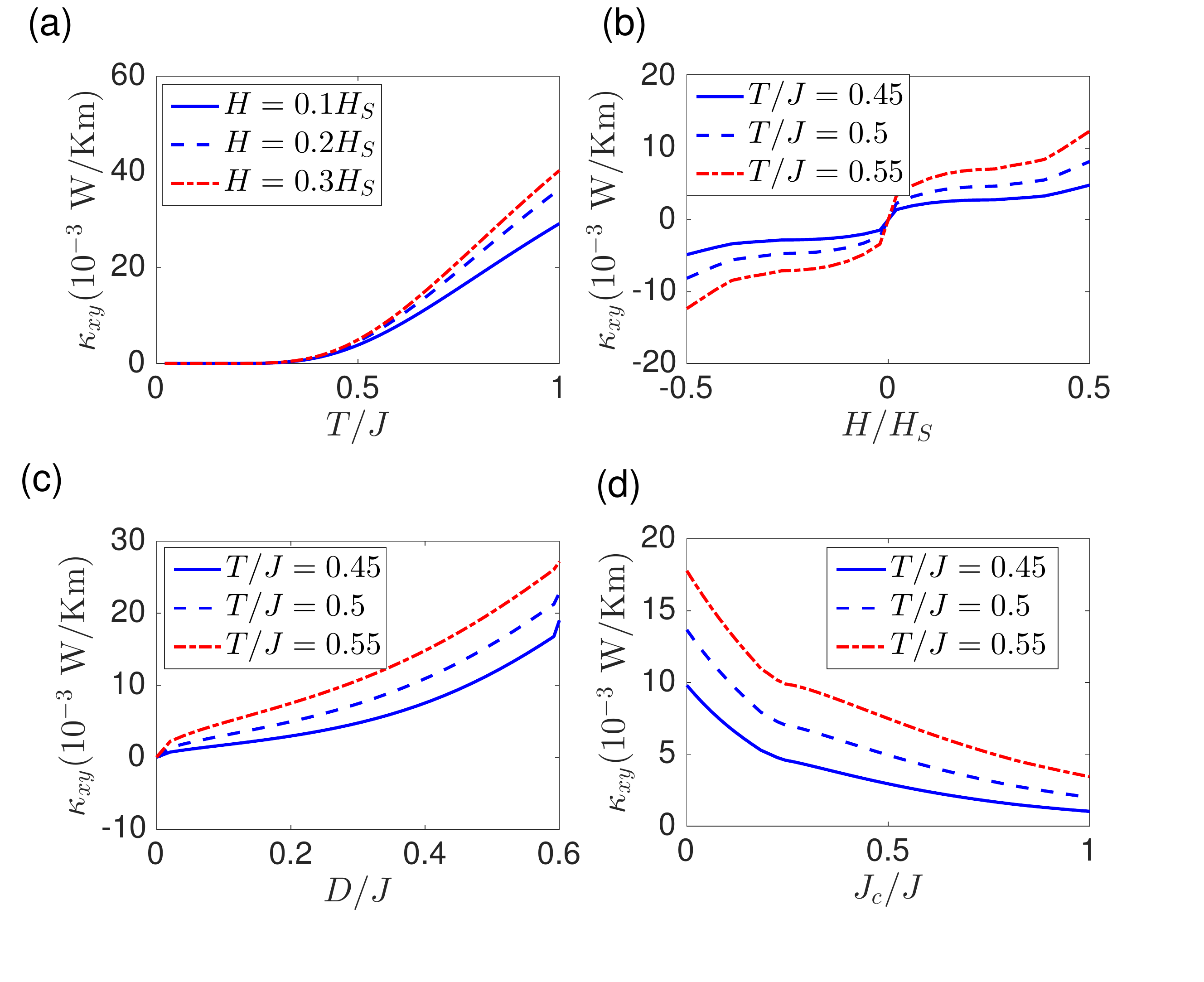}
\caption{Color online. Predicted magnon thermal Hall conductivity for layered honeycomb antiferromagnets as a function of (a) $T/J$ for different values of the magnetic field, (b) $H/H_S$, (c) $D/J$, and (d) $J_c/J$ for different values of  temperature.}
\label{THE}
\end{figure}

\section{Magnetic-field-induced magnon Chern insulator}

\subsection{Magnon Chern bands and Berry curvatures}

We will now investigate the effects of symmetry breaking in this system. We assume that the collinear magnetization lie in the $x$-$y$ plane, and a magnetic field is applied perpendicular to the plane. Hence, the spins will cant slightly along the magnetic-field direction yielding the canting angle  $ \cos\chi= H/H_S$, where the saturation field is given by $H_S = 6J + 4J_c$. Therefore, time-reversal symmetry and hence $\mathcal{PT}$ symmetry are broken.   The magnon excitations can be captured using spin wave theory as shown explicitly in the Supplemental Material.

Figure~\eqref{MCI} shows the magnon bands in the presence of a small magnetic field. We can see that the topological DNL magnons are gapped out, therefore the system transits to an antiferromagnetic magnon Chern insulator with nonzero Berry curvature and integer Chern number.  We note that the layered honeycomb-lattice non-collinear antiferromagnetic system can be considered as slices of 2D antiferromagnetic   magnon Chern insulators interpolating between the $k_z=0$ and $k_z=\pi$ planes.  Therefore for an arbitrary $k_z$ point the Chern number of the magnon bands can be defined as
\begin{align}
\mathcal C_n = \frac{1}{2\pi}\int_{BZ} d{\bf k}_\parallel\Omega^n_{xy}({\bf k}),
\end{align}
where $n$ denotes the magnon branches. In the case of nonzero magnetic field, the components of the Berry curvature  $\Omega^n_{\alpha\beta}({\bf k})$ do not vanish due to broken time-reversal symmetry. In Figs.~\ref{BC}(a) and (b) we have shown the components of the Berry curvature of the lowest magnon band on the $k_z =0$ and $k_y=0$  planes respectively. We can see that the integration of the distribution of  $\Omega^n_{xy}({\bf k})$ over the BZ  gives non-vanishing contributions as a small field-induced magnetization is along the $z$ direction. However,  the integration of the distribution of $\Omega^n_{xz}({\bf k})\equiv \Omega^n_{yz}({\bf k})$  over the BZ gives vanishing contributions  since there is no field-induced magnetization along the $x$ or $y$ direction.   

For an arbitrary $k_z$ point the Chern number is well-defined in the non-collinear spin structure.   The Chern number of all the 2D slices for an arbitrary $k_z$ point is the same, because the planes at two $k_z$ points can be adiabatically connected without closing the gap.  We find that $\mathcal C =\mp 1$ for the lowest and upper magnon bands respectively.  The presence of integer Chern numbers implies that chiral magnon edge modes transverse the bulk gap. Indeed, we  have shown in Fig.~\ref{Edge} that this is the case in the present system.

\subsection{Magnon thermal Hall effect}

In this section, we will compute the   anomalous thermal Hall conductivity of magnons \cite{th1,th2,th3,th4,th5} for the layered honeycomb antiferromagnets. In the 3D model, the total intrinsic anomalous thermal Hall conductivity has three contributions  $\kappa_{yz}, \kappa_{zx},$ and $\kappa_{xy}$, where
the components are  given by \cite{th5} 
\begin{align}
\kappa_{\alpha\beta}=- k_BT\int_{BZ} \frac{d\bo}{(2\pi)^3}~ \sum_{n=1}^N c_2[ f_n^{BE}(\bo)]\Omega_{\alpha\beta}^{n}(\bo),
\label{thm}
\end{align}
where   $ f_n^B(\bo)=\big( e^{E_{n}(\bo)/k_BT}-1\big)^{-1}$ is the Bose-Einstein distribution function, $k_B$ the Boltzmann constant which we will  set to unity, $T$ is the temperature  and $ c_2[x]=(1+x)\lb \ln \frac{1+x}{x}\rb^2-(\ln x)^2-2\text{Li}_2(-x)$, with $\text{Li}_2(x)$ being the  dilogarithm.

As the non-collinear spin configuration is induced along the $z$ direction, the first two components $\kappa_{yz}$ and $ \kappa_{zx}$ vanish. This is evident from the distribution of the Berry curvature $\Omega_{xz}^{n}(\bo)\equiv \Omega_{yz}^{n}(\bo)$ as  depicted in Fig.~\ref{BC}(b). The nonvanishing component $\kappa_{xy}$ can be written as 
\begin{align}
\kappa_{xy}= \int_{-\pi}^{\pi} \frac{d k_z}{2\pi}\kappa_{xy}^{2D}(k_z),
\label{ATHC}
\end{align}
  where $\kappa_{xy}^{\text{2D}}(k_z)$ is 2D thermal Hall conductivity for each slice of the $k_z$ plane, which is given by
 \begin{align}
\kappa_{xy}^{2D}(k_z)= -k_BT\int_{BZ} \frac{d\bo_\parallel}{(2\pi)^2}\sum_{n=1}^N c_2[ f_n^{BE}(\bo)]\Omega_{xy}^{n}(\bo_\parallel, k_z).
\label{ATHC1}
\end{align}

In Fig.~\eqref{THE} we have shown several trends of the  anomalous magnon thermal Hall conductivity. We can see that $\kappa_{xy}$ vanishes at zero temperature as no magnons are thermally excited. It also vanishes at zero magnetic field and zero DM interaction due to the presence of symmetry-protected DNL. However, a nonzero magnetic field and a nonzero  DM interaction increase $\kappa_{xy}$, whereas the interlayer coupling decreases $\kappa_{xy}$.   It is important to note that the dominant contribution to $\kappa_{xy}$ comes from the lowest  magnon branch.

\section{Conclusion}
 We have shown that layered honeycomb collinear antiferromagnets  behave differently from strictly 2D honeycomb collinear antiferromagnets. In the latter, the DM interaction-induced nonzero Berry curvature leads to a finite magnon spin Nernst response \cite{ran,kov}. In the former, however, the magnon spin Nernst effect in  each 2D antiferromagnetic layer cancels out the succeeding layer, and the system exhibits topological Dirac nodal line magnons with $\pm \pi$ Berry phase. This result implies that the antiferromagnetic interlayer coupling which always exist in the bulk insulating antiferromagnets will put a restriction  on the observation of the magnon spin Nernst response in the bulk honeycomb collinear antiferromagnets. We further showed that a magnetic-field-induced topological phase transition to a 3D antiferromagnetic magnon Chern insulator occurred in this system, by breaking the symmetry protection of  the topological Dirac nodal line magnons. We showed that topologically-protected Chern magnon bands with nonzero Berry curvatures  are manifested in the 3D antiferromagnetic magnon Chern insulator phase. We also predicted that the magnon thermal Hall effect can be observed  in layered honeycomb antiferromagnetic materials \cite{matt,oku,doi,shia, shi1, zhou, will,kly,mcn} upon the application of an external magnetic field.  Since the layered honeycomb antiferromagnets are endowed with only two magnon branches, the topological DNLs and the magnon Chern insulator  can be clearly seen at the acoustic (lowest) magnon branch, which carries the dominant thermal Hall transport contribution due to the population effect.

\acknowledgments

Research at Perimeter Institute is supported by the Government of Canada through Industry Canada and by the Province of Ontario through the Ministry of Research
and Innovation.


\begin{thebibliography}{99}
\bibitem{note1}
For  honeycomb antiferromagnetic layers coupled ferromagnetically, there is only one single doubly-degenerate magnon band in analogy to 2D  honeycomb antiferromagnet. This system does not possess any magnon band crossing such as DNL magnonic phase.
\end{thebibliography}
\end{document}


\date{\today}
\title{Supplemental Material: Topological nodal-line magnonic phase  in layered honeycomb antiferromagnets
Supplemental Material}
\author{S. A. Owerre}
\affiliation{Perimeter Institute for Theoretical Physics, 31 Caroline St. N., Waterloo, Ontario N2L 2Y5, Canada.}

\maketitle

\section{Spin transformation} 
In the low temperature regime, the underlying spin excitations of the Hamiltonian  are spin waves ({\it i.e}~~magnons). Hence, we will perform spin wave theory. First,  let us express the spins in terms of local axes, such that the $z$-axis coincides with the spin direction. This can be done by performing a local rotation about the $z$-axis  by the spin orientated angles $\theta_{i,\ell} = 0$ or $\pi$. The rotation matrix is given by 
\begin{align}
  \mathcal{R}_z(\theta_{i,\ell})=\begin{pmatrix}
\cos\theta_{i,\ell} & -\sin\theta_{i,\ell} & 0\\
\sin\theta_{i,\ell} & \cos\theta_{i,\ell} &0\\
0 & 0 &1
\end{pmatrix}.
\end{align}
The  external magnetic field induces canting out-of-plane. Therefore, we perform another rotation about $y$-axis by the  angle $\chi$. The rotation matrix is given by 

\begin{align}
  \mathcal{R}_y(\chi)=\begin{pmatrix}
\cos\chi &0& \sin\chi\\
0 & 1 &0\\
-\sin\chi &0 & \cos\chi
\end{pmatrix}.
\end{align}

 Now, the spins transform as \begin{align}
{\bf S}_{i,\ell} =\mathcal{R}_z(\theta_{i,\ell})\cdot\mathcal{R}_y(\chi)\cdot\tilde{\bf S}_{i,\ell}
 \end{align}
 The classical ground state energy  is given by \begin{align}
E_{\text{cl}}&= \mathcal{N}S^2\Big[\frac{3J}{2}\cos 2\chi -2J_c(1-2\cos^2\chi)-H\cos\chi\Big],
\end{align}
where $\mathcal{N} = 2N$ is the total number of sites and $N$ is the number of sites per unit cell. The magnetic field has been rescaled in units of $S$.  Minimizing this energy yields the canting angle $\cos\chi = {H}/{H_S}$, where $H_S=6J+ 4J_c$ is the saturation field.

\section{Spin wave theory}
The terms that contribute to linear spin wave theory are  given by

  \begin{align}
  \mathcal H_J&= J\sum_{  ij,\ell}\Big[\cos\theta_{ij,\ell} \tilde{\bf S}_{i,\ell}\cdot \tilde{\bf S}_{j,\ell}\nonumber\\& +2\sin^2\lb\frac{\theta_{ij,\ell}}{2}\rb\lb\sin^2\chi  \tilde{S}_{i,\ell}^{ x}\tilde{S}_{j,\ell}^{ x} +\cos^2\chi \tilde{S}_{i,\ell}^{ z} \tilde{S}_{j,\ell}^{ z}\rb\Big],
 \end{align}
 Due to canting induced by the magnetic field, there are two contributions to the DM interaction. They are given by  
    
  \begin{align}   
\mathcal   H_{D}^{\parallel}&= D_\parallel\sum_{\la\la ij\ra\ra,\ell}\nu_{ij}\cos\theta_{ij,\ell}~ {\bf \hat z}\cdot \big( \tilde{\bf S}_{i,\ell}\times\tilde{\bf S}_{j,\ell}\big),
\end{align} 

  \begin{align}   
\mathcal   H_{D}^{\perp}&= D_\perp\sum_{\la\la ij\ra\ra,\ell}\cos\theta_{ij,\ell}~ {\bf \hat z}\cdot \big( \tilde{\bf S}_{i,\ell}\times\tilde{\bf S}_{j,\ell}\big),
\end{align} 
where  $D_\parallel=D\cos\chi$ and $D_\perp=D\sin\chi$ are the components of the DM interaction parallel and perpendicular to the magnetic field respectively. The interlayer coupling and the magnetic field transform as
\begin{align}
 \mathcal H_{J_c}&= J_c\sum_{ i,\la \ell\ell^\prime\ra}\big[\cos\theta_{\ell\ell^\prime} \tilde{\bf S}_{i,\ell}\cdot \tilde{\bf S}_{i,\ell^\prime}\nonumber\\& +2\sin^2\lb\frac{\theta_{\ell\ell^\prime}}{2}\rb\lb\sin^2\chi  \tilde{S}_{i,\ell}^{ x}\tilde{S}_{i,\ell^\prime}^{ x} +\cos^2\chi \tilde{S}_{i,\ell}^{ z} \tilde{S}_{i,\ell^\prime}^{ z}\rb\big],\\\mathcal H_Z &= -H\cos\chi\sum_{i,\ell} \tilde{S}_{i,\ell}^{ z},
  \end{align}
where $\theta_{\alpha\beta}=\theta_{\alpha}-\theta_{\beta}=\pi$.

Next, we perform the linearized Holstein Primakoff (HP)  transformation 
\begin{align}
\tilde{S}_{i,\ell}^{ z}= S-a_{i,\ell}^\dagger a_{i,\ell},~\tilde{S}_{i,\ell}^+\approx \sqrt{2S}a_{i,\ell}=(\tilde{S}_{i,\ell}^-)^\dg,
\end{align}
where $a_{i,\ell}^\dagger (a_{i,\ell})$ are the bosonic creation (annihilation) operators, and  $\tilde{S}^\pm_{i,\ell}= \tilde{S}^x_{i,\ell} \pm i \tilde{S}^y_{i,\ell}$ denote the spin raising and lowering  operators.  The resulting noninteracting magnon Hamiltonian can be written as
\begin{widetext}
\begin{align}
\mathcal H&= JS\sum_{\la ij\ra,\ell}\big[-t_{\chi}(a_{i,\ell}^\dg a_{j,\ell} +{\rm H.c.})- (1-t_{\chi})(a_{i,\ell}^\dg a_{j,\ell}^\dg + {\rm H.c.})\big]+t_D^\parallel S\sum_{\la\la ij\ra\ra,\ell} (e^{-i\nu_{ij}\phi}a_{i,\ell}^\dg a_{j,\ell} + {\rm H.c.}) \nonumber\\&+t_D^\perp S\sum_{\la\la ij\ra\ra,\ell} (e^{i\phi}a_{i,\ell}^\dg a_{j,\ell} + {\rm H.c.})- (1-t_{\chi})(a_{i,\ell}^\dg a_{j,\ell}^\dg + {\rm H.c.})\big]\nonumber\\&+ J_cS\sum_{i,\la \ell\ell^\prime\ra}\big[ - t_\chi (a_{i,\ell}^\dg a_{i,\ell^\prime} + {\rm H.c.}) + (1-t_{\chi})(a_{i,\ell}^\dg a_{i,\ell^\prime}^\dg + {\rm H.c.})\big]+ t_0S\sum_{i,\ell}a_{i,\ell}^\dg a_{i,\ell},
\end{align}
\end{widetext}
where
\begin{align}
&\phi  = \pi/2, ~t_0 = 3JS+2J_cS, \\& t_{\chi}= \cos^2\chi,~t_D^\parallel = D\cos\chi,~t_D^\perp =D\sin\chi. 
\end{align}
In momentum space, we obtain  
 \begin{align}
 \mathcal H=\frac{1}{2}S\sum_{\bf k} \psi^\dg({\bf k})\cdot\mathcal H({\bf k})\cdot\psi({\bf k}),
 \end{align}
where
\begin{align}
 \mathcal H({\bf k})=
 \begin{pmatrix}
\mathcal{A}({\bf k})&  \mathcal{B}({\bf k})\\  \mathcal{B}^*(-{\bf k})&  \mathcal{A}^*(-{\bf k})
 \end{pmatrix}.
 \label{hamp}
\end{align}
where
\begin{widetext}
\begin{align}
\mathcal{A}({\bf k})&=
 \begin{pmatrix}
f(k_z)-m^\parallel({\bf k}_\parallel)+m^\perp({\bf k}_\parallel)&-t_{\chi}f({\bf k}_\parallel)\\ -t_{\chi}f^*({\bf k}_\parallel) &f(k_z)+m^\parallel({\bf k}_\parallel)+m^\perp({\bf k}_\parallel)
 \end{pmatrix},\quad  \mathcal{B}({\bf k})&=
 \begin{pmatrix}
\tilde{f}(k_z)&(1-t_{\chi})f({\bf k}_\parallel)\\ (1-t_{\chi})f^*({\bf k}_\parallel) &\tilde{f}(k_z)
 \end{pmatrix},
\end{align}
\end{widetext}

with

\begin{align}
&f(k_z)=t_0 - 2J_c t_\chi \cos k_z,\\&\tilde{f}(k_z)= 2J_c (1-t_{\chi}) \cos k_z,\\& f({\bf k}_\parallel) = J\big(1 + e^{-ik_{\parallel}^1}+e^{-ik_{\parallel}^2}\big),\\& 
m^{\parallel(\perp)}({\bf k}_\parallel)\nonumber\\& = 2D\cos\chi(\sin\chi)\big[\sin(k_{\parallel}^1) -\sin(k_{\parallel}^2) +\sin\big(k_{\parallel}^1-k_{\parallel}^2\big)\big].
\end{align}

\section{Paraunitary operators}
The magnon Hamiltonian $\mathcal{H}(\bo)$  can be  diagonalized using the bosonic Bogoliubov transformation:  $\psi_\bo= \mathcal U_\bo Q_\bo$, 
where $ Q_\bo^\dg=(\gamma_{\bo A}^{\dg}, \gamma_{\bo B}^{\dg}, \gamma_{-\bo A}, \gamma_{-\bo B})$ are the Bogoliubov quasiparticles, and $\mathcal U_\bo$ is a $2N\times 2N$ paraunitary matrix  defined as
\begin{align}
& \mathcal U_\bo= \begin{pmatrix}
  u_\bo & v_\bo \\
v_\bo & u_\bo\\  
 \end{pmatrix},
\end{align} 
where $u_\bo$ and $v_\bo$ are  $N\times N$ matrices that satisfy \bea |u_\bo|^2-|v_\bo|^2= {\rm {\bf I}}_{N\times N}.\eea  The matrix $\mathcal U_\bo$ satisfies the relations,
\begin{align}
 &\mathcal U_\bo^\dg  \tau_3 \mathcal U_\bo=  \tau_3, \label{eqn1}~
  \mathcal U_\bo^\dg \mathcal{H}(\bo) \mathcal U_\bo=\mathcal{E}(\bo),
\end{align}
where
\begin{align}
\mathcal{E}(\bo) = \begin{pmatrix}E_{n}(\bo) & 0\\ 0 & E_{n}(-\bo)\end{pmatrix},~  \tau_3=\begin{pmatrix}{\rm {\bf I}}_{N\times N} & 0\\ 0 & -{\rm {\bf I}}_{N\times N}\end{pmatrix},
\end{align}
with    energy band $E_{n}(\bo)$ and band index  $n$.  The bosonic Bogoliubov Hamiltonian is given by $\mathcal{H}_B(\bo)=  \tau_3\mathcal{H}(\bo)$,  whose eigenvalues are given by $  \tau_3\mathcal E_\bo$ and the columns of $\mathcal U_\bo$ are the corresponding eigenvectors.

At zero magnetic field $H=0$, we find that  the   $N\times N$ matrices (with $N =2$) $u_\bo$ and $v_\bo$ that satisfy the relations are  given by
\begin{align}
u_\bo=\frac{1}{\sqrt{2}}
\begin{pmatrix}
-e^{i\varphi_{{\bf k_\parallel}}}\cosh\phi_\bo & e^{i\varphi_{{\bf k_\parallel}}}\cosh\theta_\bo\\
-\cosh\phi_\bo & -\cosh\theta_\bo
\end{pmatrix},
\end{align}

\begin{align}
v_\bo=\frac{1}{\sqrt{2}}
\begin{pmatrix}
e^{i\varphi_{{\bf k_\parallel}}}\sinh\phi_\bo & -e^{i\varphi_{{\bf k_\parallel}}}\sinh\theta_\bo\\
\sinh\phi_\bo & \sinh\theta_\bo
\end{pmatrix},
\end{align}
where 
\begin{align}
&\tanh 2\phi_\bo=\frac{f(k_z)+f(\bo_\parallel)}{f_0},\label{tan1}\\&
\tanh 2\theta_\bo=\frac{f(k_z)-f(\bo_\parallel)}{f_0}\label{tan2},\\&
\tan\varphi_{{\bf k_\parallel}}=\frac{\mathfrak{I}[f({\bf k_\parallel})]}{\mathfrak{R}[f({\bf k_\parallel})]},
\end{align}
where $\mathfrak{R}$ and $\mathfrak{I}$ denote real and imaginary parts.
 For nonzero magnetic field $H\neq 0$, we have diagonalized the Hamiltonian numerically.